\documentclass[%
 reprint,
 amsmath,amssymb,
 aps,
 prl,
]{revtex4-2}

\usepackage{flushend}
\usepackage{verbatim}
\usepackage{verbatim,upgreek}
\usepackage{chngcntr}
\usepackage{graphicx}
\usepackage{dcolumn}
\usepackage{bm}
\usepackage{comment}
\usepackage{amsmath}
\usepackage{physics}
\usepackage{hyperref}
\hypersetup{
    colorlinks=true,
    citecolor=blue,
    }

\usepackage{color}

\newcommand{\ms}[1]{\mbox{\scriptsize #1}}

\begin{document}

\preprint{APS/123-QED}

\title{Dispersive readout of cavity-coupled solid-state sensor with near-unity readout fidelity}

\author{Hanfeng Wang$^{1,4}$}
\email{hanfengw@mit.edu}

\author{Shuang Wu$^{1,2,4}$}


\author{Matthew E. Trusheim$^{1,3}$}

\author{Avetik Harutyunyan$^{1,2}$}

\author{Dirk R. Englund$^{1}$}

\affiliation{\\
$^{1}$ Massachusetts Institute of Technology, 50 Vassar Street, Cambridge, MA 02139, USA\\
$^{2}$ Honda Research Institute USA, Inc., San Jose, CA 95134, USA\\
$^{3}$ U.S. Army DEVCOM Army Research Laboratory, Adelphi, MD 20783, USA\\
$^{4}$ These authors contributed equally to this work.}

\begin{abstract}

Solid-state quantum sensors based on ensembles of nitrogen-vacancy (NV) centers in diamond have emerged as powerful platforms for high-precision metrology. Coupling the NV ensemble to a microwave cavity mode in a cavity quantum electrodynamics (cQED) configuration enables spin readout that surpasses the limitations of conventional optical detection, achieving sub-picotesla magnetic sensitivities. However, existing continuous-wave cQED approaches remain far from the intrinsic spin-projection-noise limit due to spin saturation and power broadening.
Here, we introduce a dispersive cQED readout technique to overcome these fundamental limitations in NV ensemble sensing. We develop a comprehensive theoretical framework describing the dispersive interaction and analyze the time-domain dynamics of a strongly-coupled NV-cavity system. Our results indicate near-unity inverse readout fidelity and femtotesla-level sensitivity using a commercially available diamond NV ensemble. Importantly, the dispersive readout exhibits a distinct sensitivity scaling that improves as $1/N$ with increasing number of spins $N$, providing a practical pathway toward approaching the standard quantum limit for solid-state spin-ensemble sensors.

\end{abstract}

\maketitle

\textit{Introduction - } Quantum sensing has emerged as a transformative technology capable of achieving unprecedented sensitivity, approaching fundamental quantum-mechanical limits \cite{degen2017quantum,han2026ultra,du2024single}. Among various quantum sensing platforms, nitrogen-vacancy centers (NV) in diamond are particularly compelling, exhibiting favorable properties such as nanoscale spatial resolution \cite{pelliccione2016scanned,kim2023nanophotonic,wang2020electrical}, robust spin coherence at room temperature \cite{bauch2018ultralong,stanwix2010coherence}, and intrinsic compatibility with harsh environments \cite{fu2020sensitive}. These distinct characteristics position NV-based quantum sensors as versatile candidates for high-performance magnetometry \cite{wolf2015subpicotesla,fescenko2020diamond}, electrometry \cite{dolde2011electric,wang2023field}, and inertial sensing applications \cite{soshenko2021nuclear,jarmola2021demonstration}.
Unlike traditional sensitive magnetometry methods, such as superconducting quantum interference devices (SQUIDs) and spin-exchange relaxation-free (SERF) atomic sensors \cite{griffith2010femtotesla,trabaldo2020squid}, NV-based sensors do not require stringent operating conditions like cryogenic cooling or extensive magnetic shielding. However, despite these advantages, practical NV sensors generally exhibit sensitivities limited to the picotesla regime \cite{barry2020sensitivity,eisenach2021cavity,wolf2015subpicotesla,fescenko2020diamond}, significantly above their fundamental sensitivity bounds.
The fundamental sensitivity limit of spin-based quantum sensors, known as the standard quantum limit (SQL), is defined as $\eta = 1/\gamma_{\mathrm{e}}\sqrt{NT_2^*}$, involving the gyromagnetic ratio $\gamma_{\mathrm{e}}$, coherence time $T_2^*$, and total number of spins $N$. Single NV centers can operate close to the SQL ($\sigma_{\mathrm{e}} \sim 1$) but are intrinsically restricted to nano-tesla sensitivity levels due to the limited spin number \cite{bonato2016optimized}. In contrast, ensemble-based NV sensors leverage large spin numbers ($\sqrt{N} \sim 10^6$), but optical readout leads to large inverse readout fidelities due to low photon detection efficiency (typically $\sigma_{\mathrm{e}} \sim 10^3$ in state-of-the-art experiments), ultimately limiting sensitivity improvements \cite{schloss2018simultaneous,barry2016optical,barry2020sensitivity}.

Recently, cavity-enhanced NV spin sensors have demonstrated significant progress, achieving inverse readout fidelities as low as $\sigma_e \sim 360$, approaching the SQL \cite{wang2024spin,wang2025cavity,wang2025strongly,eisenach2021cavity,wang2026ultralow}. Further enhancements have been reported using phase-transition-based sensing, reaching $\sigma_e \sim 100$ \cite{wang2026exceptional}.
However, conventional continuous-wave cavity measurement techniques suffer from microwave-induced power broadening and spin depolarization, which reduce the effective coupling strength between the NV ensemble and the cavity mode \cite{wang2024spin,zhang2021exceptional,angerer2017ultralong,scully1999quantum}, thereby limiting further improvement in $\sigma_e$.
Pulsed measurement protocols provide a natural route to overcome these limitations. By temporally separating spin manipulation and readout, pulsed sequences reduce microwave exposure, thereby suppressing power broadening and preserving spin polarization. In addition, dynamical decoupling and spin-echo-based techniques can extend coherence times by mitigating environmental noise and inhomogeneous broadening \cite{barry2020sensitivity,barry2024sensitive}. Although such pulsed protocols are widely used and form the standard approach in optical NV sensing, their implementation in microwave cavity-based readout remains largely unexplored.


\begin{figure}[t!]
    \includegraphics[width = 0.48\textwidth]{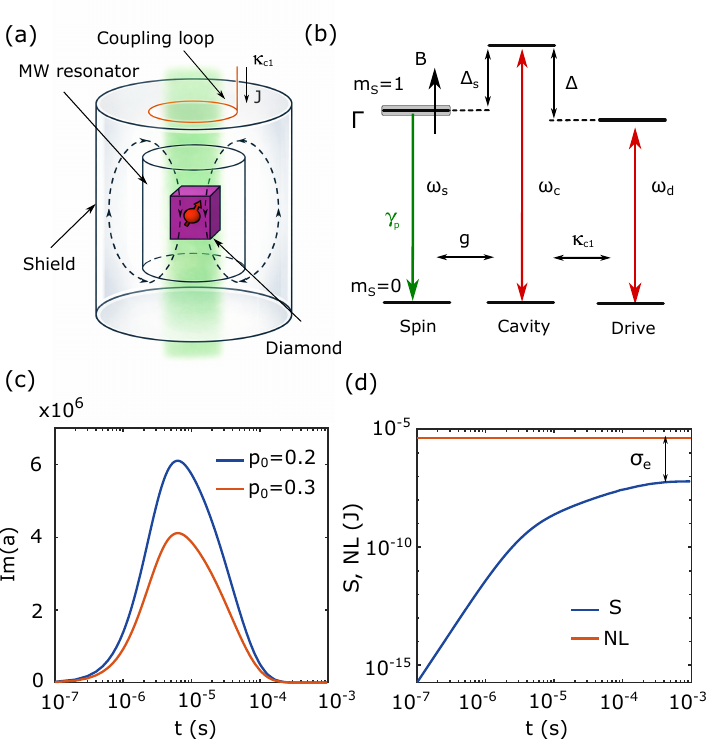}
    \caption{\textbf{NV-cQED dispersive readout.} (a) Schematic of the NV-cQED system. A diamond containing an NV ensemble is placed at the center of a dielectric resonator. The cavity field $\alpha$ is detected via a coupling loop with an external coupling rate $\kappa_{c1}$. The system is enclosed within a microwave shield. (b) NV–cavity energy-level structure. (c) Time dynamics of the cavity quadrature $\mathrm{Im}[\alpha]$ for initial excited-state populations $p = 0.2$ (blue) and $p = 0.3$ (red). (d) Time dynamics of signal $S$ (blue) and the product $N\mathcal{L}$ (red) as a function. The ratio of noise and signal is defined as the inverse readout fidelity $\sigma_e$. }
    \label{fig1}
\end{figure}

In pulsed sensing schemes, on-resonant spin-cavity interactions introduce two key limitations. First, during the evolution period, the probe field drives real spin transitions, leading to population transfer and spontaneous emission into the cavity mode. This process disrupts the coherent accumulation of phase induced by the target field. Second, the readout field itself perturbs the spin population, inducing fluctuations and effectively destroying the prepared quantum state during measurement. These effects fundamentally undermine the advantages of pulsed sensing, where preserving coherence throughout the measurement sequence is essential.
Dispersive readout offers a natural pathway to overcome these limitations. By operating in the off-resonant regime, spin states are inferred through cavity frequency or phase shifts without directly driving the spin transition, thereby suppressing measurement destruction and preserving coherence throughout the sensing sequence. In hybrid quantum architectures such as superconducting qubits and cold atomic ensembles \cite{crippa2019gate,kohler2017dispersive,peronnin2020sequential,park2020adiabatic}, dispersive readout has become the standard approach for high-fidelity state detection.
However, when extended to NV ensemble systems, existing theoretical treatments \cite{ebel2021dispersive,kozodaev2025dispersive} typically describe the accumulated phase shift solely in terms of the initial and final states, effectively treating the measurement as a static process and neglecting the full time dynamics of the coupled spin–cavity system. 
As a result, these approaches do not capture the interplay between the microwave cavity mode and the spin ensemble during the measurement. 
This omission becomes particularly important in thermally limited, room-temperature systems, where the finite measurement time motivates stronger driving to enhance the signal. In this regime, the noise is largely independent of the drive, so increasing power directly improves the readout fidelity until limited by measurement-induced depolarization, requiring a full time-dependent treatment of the system dynamics. Moreover, for sensing applications, the overall sensitivity performance not only on readout fidelity, but also depends on the full measurement cycle, including initialization and readout overhead, further necessitating a time-dependent description.





In this work, we develop a comprehensive Maxwell-Bloch framework for NV-based cavity quantum electrodynamic (NV-cQED) systems that captures the full dynamical evolution of the coupled spin–cavity system. Our analysis reveals that, once the phase noise of the probe is taken into account, optimal performance can emerge in an intermediate regime with moderate detuning, beyond the conventional deeply dispersive limit. We demonstrate that the current setting achieves near-unity inverse readout fidelity, corresponding to more than an order-of-magnitude improvement over conventional continuous-wave NV-cQED approaches \cite{wang2024spin}, together with femtotesla-level sensitivity. Notably, the dispersive readout exhibits a distinct sensitivity scaling that improves as $1/N$, where $N$ is the number of spins. This provides a viable route to overcome the trade-off between ensemble size and readout fidelity, enabling scalable quantum sensors to approach the SQL in solid-state spin-ensemble systems.

\textit{Cavity-coupled NV system} - We consider the interaction of an NV system with a quantized cavity field $a$ coupled to the spin transition $|0\rangle\leftrightarrow|1\rangle$ and driven by an external probe field $J$, as shown in Fig. 1(a). The Hamiltonian of the coupled NV-cavity system reads \cite{scully1999quantum}:
\begin{equation}
\mathcal{H}=\Delta a^{\dagger} a + iJ(a^{\dagger}-a) +\sum_{j=1}^N\Delta_{\ms{s}}^{(j)} \sigma_{j}+ \sum_{j=1}^N g_{\ms{s}}^{(j)} (a \sigma_{j}^\dagger+a^{\dagger} \sigma_{j}) 
\end{equation}
In the following analysis, we assume a homogeneous spin ensemble and therefore omit the spin index $j$. This approximation is justified in the dispersive regime \cite{Supplementary}. The operator $a$ denotes the annihilation operator of the cavity mode. The operator $\sigma$ denotes the lowering operator of the NV spin, and $g_{\mathrm{s}}$ is the single-photon coupling strength; the collective coupling strength of the ensemble is given by $g = \sqrt{N} g_{\mathrm{s}}$. The detunings are defined as $\Delta_c = \omega_c - \omega_d$ for the cavity-probe detuning and $\Delta_s = \omega_s - \omega_d$ for the spin-probe detuning, where $\omega_d$, $\omega_c$, and $\omega_s$ denote the drive, cavity, and spin frequencies, respectively, as illustrated in Fig.~1(b). 



\begin{figure}[!]
    \includegraphics[width = 0.47\textwidth]{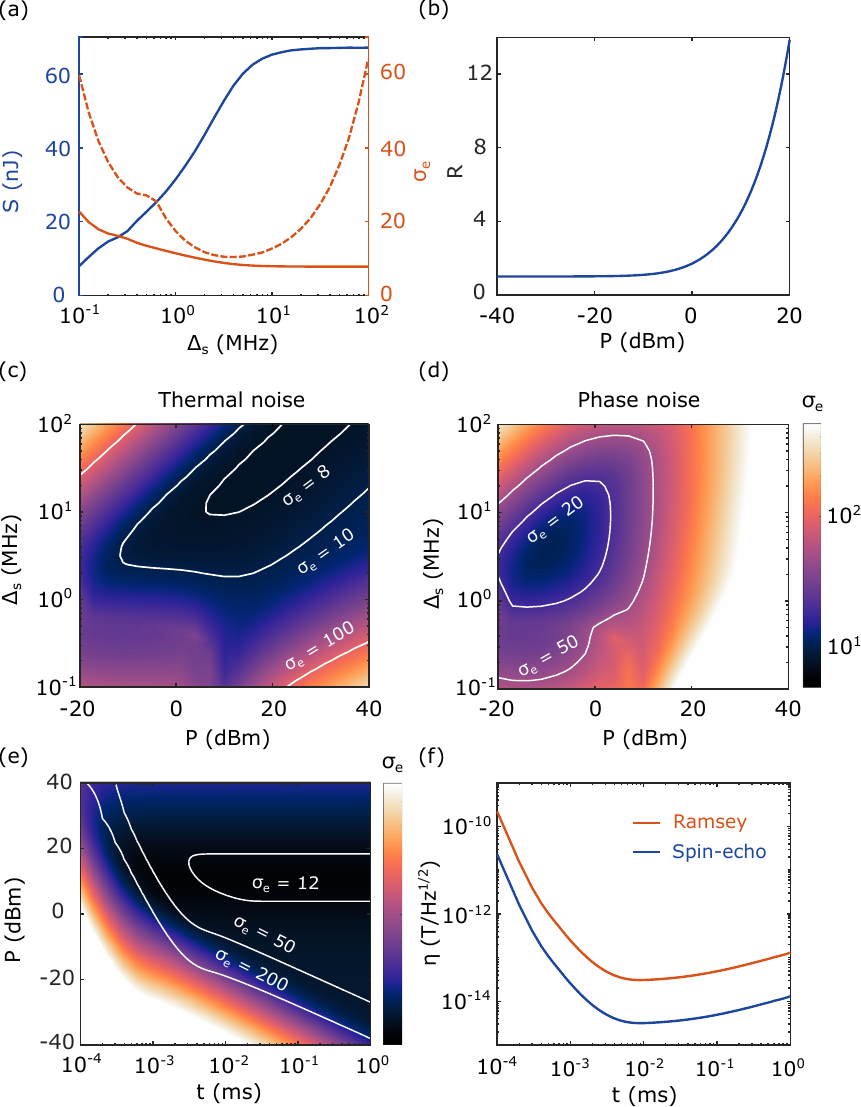}
    \caption{\textbf{Inverse readout fidelity and sensitivity performance.} 
(a) Signal $S$ (blue) and inverse readout fidelity $\sigma_e$ (red) as functions of the spin–cavity detuning $\Delta_s$. Simulation parameters: $\kappa_c = 2\pi \times 130~\mathrm{kHz}$, $\kappa_{c1} = 2\pi \times 130~\mathrm{kHz}$, $g_s = 2\pi \times 0.019~\mathrm{Hz}$, $N = 10^{15}$, and $\Gamma = 2\pi \times 330~\mathrm{kHz}$. Time-domain simulations are employed to capture the regime $\Delta_s \sim \Gamma$. Those parameters are accessible with experiments \cite{wang2024spin}.
Solid line: $\sigma_e$ limited by thermal noise. Dashed line: $\sigma_e$ including phase noise.
(b) Ratio $R$ between phase noise and thermal noise as a function of input power $P$.
(c,d) $\sigma_e$ as a function of input power $P$ and detuning $\Delta_s$ with fixed measurement time $t = 1$ ms limited by thermal noise (c) or phase noise (d).
(e) $\sigma_e$ as a function of input power $P$ and measurement time $t$ at $\Delta_s = 2~\mathrm{MHz}$. 
(f) Sensitivity as a function of measurement time $t$ for Ramsey (red) and spin-echo (blue) sequences, limited by $T_2^*$ and $T_2$, respectively.
}
    \label{fig3}
\end{figure}

We employ the quantum master equation to solve the cavity mode dynamics based on the Hamiltonian in Eq. (1), incorporating loss processes within the Lindbladian open-system formalism \cite{wang2024spin}. The cavity mode undergoes relaxation at a total rate of $\kappa = \kappa_{\mathrm{c}} + \kappa_{\mathrm{c}1}$, where $\kappa_{\mathrm{c}}$ represents intrinsic losses and $\kappa_{\mathrm{c}1}$ accounts for coupling to a microwave loop line. The NV ensemble is subject to depolarization and decoherence processes:  electron spin decoherence at rate $\Gamma$; thermal depolarization at rate $\gamma_{\mathrm{th}}$. 

\textit{Analytical solution in the dispersive regime - }We provide an analytical solution for the Maxwell-Bloch equations governed by the Hamiltonian in Eq. (1). We consider the system in the dispersive regime, where $\Delta_s \gg \kappa, \Gamma$. In this regime, the characteristic timescale for the spin coherence $\sigma$ is given by $\tau_\sigma \sim 1/|\Delta_s|$. For large detunings $|\Delta_s|$, this timescale becomes very short, causing $\sigma$ to rapidly approach its instantaneous equilibrium determined by the slower-evolving variables \cite{park2020adiabatic}. On longer timescales (characterized by $1/\kappa$, $1/\Gamma$, etc.), the cavity field $\alpha$ and excited state population $p$ evolve slowly, justifying the assumption $d\sigma/dt = 0$. Under this approximation, the cavity field $\alpha$ has an analytical solution:
\begin{equation}
\alpha(t) = \sqrt{\kappa_{c1}}\beta\int_0^t\frac{ A(t')}{A(t)}dt',
\end{equation}
with the integral kernel given by \cite{Supplementary}:
\begin{equation}
A(t) = \exp(-\frac{\kappa}{2}t + i\frac{g_s^2N}{\Delta_s}\int_0^t (1-2p)dt').
\end{equation}
Eq. (3) and Eq. (4) show that the cavity field is determined by a time-integrated response that depends on the evolution of the spin population. In contrast to conventional dispersive readout for thermally-limited NV ensembles \cite{ebel2021dispersive}, which is typically analyzed in the deeply dispersive limit where the cavity response is effectively static, the present formulation captures the dynamical interplay between the spin ensemble and the cavity field.
This enables us to access regimes beyond the deeply dispersive limit, 
where the cavity field exerts a destructive measurement on the spin dynamics. In this intermediate regime, the readout signal is no longer determined solely by a static frequency shift, but by the coupled evolution of the spin–cavity system. 
We plot an exemplary time dynamics of the phase of the cavity field Im($\alpha$) for initial excited state population $p_0=0.2$ and $p_0=0.3$ in Fig. 1(c).

The inverse readout fidelity for the spin ensemble with size $N$ could be expressed as $\sigma_e=\sqrt{N\mathcal{L}/S}$. The noise performance $\mathcal{L}$ is limited by thermal noise, phase noise, and other noise sources. The measured signal could be expressed as an integral of the cavity phase accumulation:
 \begin{equation}
 \begin{aligned}
     S =\hbar\omega_c\kappa_{c1} \int_0^{t_r}\frac{\mathrm{Im}^{2}[\alpha(p_0)] - \mathrm{Im}^{2}[\alpha(p_0')]}{p_0-p_0'}d\tau
  \end{aligned}
 \end{equation}
Here, $p_0$ and $p_0'$ denote the excited-state populations corresponding to the two spin states to be discriminated during the readout process. In Fig.~1(d), we plot the signal $S$ together with the product $N\mathcal{L}$, where $\mathcal{L}$ characterizes the noise. The ratio of these quantities defines the inverse readout fidelity. We observe that this ratio converges to a constant as the measurement time approaches the spin relaxation time with millisecond time scale.

We further plot the modeled signal and inverse readout fidelity as functions of the spin–cavity detuning $\Delta_s$ in Fig.~2(a), assuming the noise is limited by thermal fluctuations, $\mathcal{L_{\mathrm{th}}} = k_B T$. The number of spins is taken to be $N = 10^{15}$, limited by optical polarization under current experimental conditions \cite{wang2024spin}.
In this analysis, the inverse readout fidelity is optimized over the input power. For larger detuning $\Delta_s$, higher input power can be applied without inducing destructive measurements, thereby maintaining the signal $S$. Under this idealized assumption, where thermal noise is the sole limitation, the inverse readout fidelity reaches $\sigma_e = 8$, representing nearly two orders of magnitude improvement over conventional continuous-wave readout methods under comparable conditions \cite{wang2024spin}.

When thermal noise is the dominant limitation, the optimal $\sigma_e$ is independent of the spin-cavity detuning $\Delta_s$ in the dispersive regime, since a larger detuning can, in principle, always be compensated by increasing the drive power. In practice, however, this picture breaks down once phase noise in the input field is taken into account. Increasing the drive power introduces additional noise, in particular phase noise, which can eventually exceed the thermal noise and become the dominant contribution. The phase noise can be expressed as \cite{wang2024spin,Supplementary,wilcox2022thermally}:
\begin{equation}
\mathcal{L}_{\mathrm{ph}} = |\Phi(f)\Gamma_p(f)|^2,
\end{equation}
where $f = 1/t$ with $t$ the measurement time. Here, $\Phi(f)$ denotes the single-sideband phase noise of the input field, and $\Gamma_p(f)$ is the reflection coefficient of the NV-cavity hybrid system. As a representative example, we adopt the noise spectrum of the Rohde \& Schwarz SMA100B, which offers state-of-the-art phase noise performance among commercially available signal generators \cite{Supplementary}.

Phase noise enters the system through the input field and depends on the drive power. We plot the ratio between phase noise and thermal noise, $R \equiv \sqrt{\mathcal{L}_{\mathrm{ph}} / \mathcal{L}_{\mathrm{th}}+1}$, as a function of input power in Fig.~2(b). This ratio remains on the order of unity at low input power, but increases rapidly once $P$ exceeds 0 dBm. As shown in Fig.~2(c), the optimal $\sigma_e$ occurs in the regime with $P > 0$ dBm, where phase noise can become the dominant noise contribution.
Taking phase noise into account, the resulting $\sigma_e$ is shown as the dashed curve in Fig.~2(a), with the corresponding dependence in power–detuning space illustrated in Fig.~2(d). We find that $\sigma_e$ degrades significantly when $\Delta_s$ is either too large or too small, with an optimum emerging in the intermediate regime. Notably, this optimal performance remains close to the thermal-noise limit of the system.


\begin{figure}
    \includegraphics[width = 0.49\textwidth]{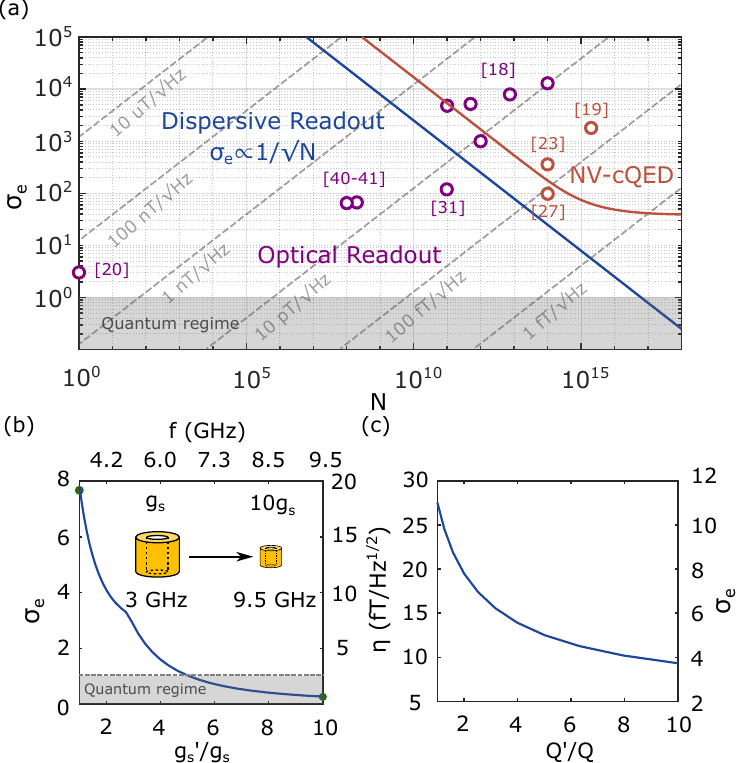}
    \caption{\textbf{Approaching SQL}. (a) The inverse readout fidelity $\sigma_e$ as a function of spin number $N$. The previous work based on the optically detected magnetic resonance is marked as purple. The continuous-wave readout initially exhibits a sensitivity scaling of $\sigma_e \propto 1/\sqrt{N}$, but gradually saturates for ensemble sizes $N > 10^{16}$. In contrast, the dispersive readout method maintains a scaling of $\sigma_e \propto 1/\sqrt{N}$, closely approaching the SQL scaling of $\sigma\sim 1$ for large spin ensembles. We indicate the quantum regime with shaded regions, as it is not captured by our model. References: \cite{le2012efficient,wang2026exceptional,wang2024spin,eisenach2021cavity,barry2020sensitivity,chatzidrosos2017miniature,bonato2016optimized} (b) Inverse readout fidelity and sensitivity as a function of resonant frequency and the coupling strength enhancement ratio. (c) Inverse readout fidelity and sensitivity as a function of the quality factor enhancement ratio.  }
    \label{fig4}
\end{figure}

\textit{Sensitivity optimization - }
As a sensor, the ultimate sensitivity depends not only on the inverse readout fidelity $\sigma_e$, but also on experimental overhead time, and can be expressed as \cite{barry2020sensitivity,degen2017quantum}:
\begin{equation}
\eta = \frac{\sigma_e}{\gamma_e \sqrt{N T}} \sqrt{\frac{t_f}{T}},
\end{equation}
where $t_f$ denotes the total experimental time, including initialization, evolution, and readout. Here, $T$ represents the characteristic sensing time, which depends on the measurement protocol and may correspond to the inhomogeneous dephasing time $T_2^*$, the coherence time $T_2$, or other relevant timescales.

As shown in Fig.~2(e), the inverse readout fidelity saturates with increasing readout time $t$. However, longer measurement durations also increase the experimental overhead, which ultimately degrades the overall sensitivity. As illustrated in Fig.~2(f), the sensitivity reaches $40~\mathrm{fT}/\sqrt{\mathrm{Hz}}$ and $4~\mathrm{fT}/\sqrt{\mathrm{Hz}}$ for Ramsey and spin-echo sequences, respectively, at intermediate measurement times for the current device. These results represent an order-of-magnitude improvement in sensitivity over conventional continuous-wave readout methods and are comparable to state-of-the-art magnetometers based on SQUID and SERF technologies. 

\textit{Inverse readout fidelity scaling - }We next examine the scaling behavior of the inverse readout fidelity. 
For optical readout, the single-spin fidelity can approach unity, thus operating near the standard quantum limit (SQL). 
However, for large ensembles ($N > 10^{11}$), the inverse readout fidelity of conventional optical detection deteriorates to values on the order of $\sigma_e \sim 5000$ (with the state-of-the-art measurement approaching $\sigma_e\sim 10^2$ for $N\sim10^{11}$ \cite{barry2023sensitive}), primarily due to limited photon collection efficiency, as shown in Fig.~3(a). 
In contrast, NV-cQED readout provides a pathway to enhance $\sigma_e$ for large spin ensembles. 
As illustrated in Fig.~3(a), the continuous-wave NV-cQED readout follows a scaling of $\sigma_e \propto 1/\sqrt{N}$, offering improved performance with increasing ensemble size. 
Nevertheless, this scaling eventually saturates for very large $N$ due to spin saturation and power broadening effects, which limit further improvement in $\sigma_e$ \cite{wang2024spin}. 

The proposed dispersive readout overcomes these limitations by operating in the non-saturating, off-resonant regime, thereby preserving the favorable scaling $\sigma_e \propto 1/\sqrt{N}$ (corresponding to $\eta \propto 1/N$, dashed line in Fig. 3(a)), achieving state-of-the-art readout fidelity, and enabling further approach to the SQL.
(1) We first compare the present system to the single-spin limit under the condition of equal total coupling strength (see Supplementary Information Sec.~IB). In the ensemble case, the collective interaction leads to an enhancement of the cavity field that scales as $\alpha \propto \sqrt{N}$ with the number of spins $N$.
(2) In the single-spin limit, the cavity field scales linearly with the coupling strength, $\alpha \propto g \propto \sqrt{N}$. Combining these results, the final cavity field will be proportional to number of spins $N$, leading to the signal scaling $S\propto N^2$ and inverse readout fidelity scales as $\sigma_e \propto 1/\sqrt{N}$.
However, we emphasize that while semiclassical models predict SQL surpassing for ensembles larger than $3\times10^{17}$ spins, quantum effects near the semiclassical-quantum transition boundary may alter this prediction \cite{krimer2019critical,zens2019bistablequantum}, therefore we marked a ``quantum regime" which is not captured in our semiclassical model in Fig. 3(a). Exploring such quantum regimes lies beyond the scope of the present study and will be explored in future work.

\textit{Conclusion and outlook - } 
In the present discussion, we restrict the number of spins to $N \sim 10^{15}$, as limited by the optical polarization efficiency of our current system \cite{Supplementary}. This constraint can, however, be relaxed through alternative cavity designs or operation at higher resonant frequencies. In particular, increasing the cavity frequency via an applied magnetic field enhances the single-spin coupling strength, as $g_s \propto f^2$. As shown in Fig.~3(b), we plot the inverse readout fidelity and sensitivity as the coupling strength is scaled from $g_s$ to $g_s' = \xi g_s$. This scaling leads to an improvement in both $\sigma_e$ and sensitivity by approximately one order of magnitude when the resonant frequency is increased from 3 GHz to 10 GHz.
It is also worth noting that $\sigma_e$ improves with increasing $\xi$ but eventually saturates at large $\xi$, rather than continuing to scale linearly with the cooperativity or the dispersive frequency shift. This behavior indicates that nonlinear saturation effects remain significant even in the dispersive readout regime when operating near the thermal-noise-limited sensitivity optimum. We further support this conclusion by analyzing the effect of increasing the cavity quality factor by $Q' = \xi Q$ in Fig. 3(c), which exhibits a similar nonlinear trend and thus confirms the presence of intrinsic nonlinear dynamics in the system.

Our proposed dispersive readout achieves near-unity inverse readout fidelity and a sensitivity of $4~\mathrm{fT}/\sqrt{\mathrm{Hz}}$, highlighting its strong potential for advancing cQED-based sensor performance. The favorable $1/N$ scaling with NV ensemble size provides a promising pathway toward approaching the SQL.
Future work will explore several directions guided by this dispersive framework, including a fully quantum mechanical treatment of large NV ensembles, systematic comparisons with dispersive readout schemes based on single superconducting qubits through scaling of the collective coupling strength, and the experimental realization of NV-cQED sensors. While the present discussion focuses on magnetometry, this readout approach is broadly applicable to other sensing modalities, including gyroscopes \cite{wang2025cavity}, atomic clocks \cite{trusheim2020polariton}, and electric-field sensing \cite{dolde2011electric}.

\textit{Acknowledgment - } The authors would like to thank John Rich and Bejoy Sidker for helpful discussions. This work is supported by Honda Research Institute USA. D.R.E.~acknowledges funding from the MITRE Corporation and the U.S.~NSF Center for Ultracold Atoms.

\bibliography{apssamp}

\end{document}